\documentclass[preprint,preprintnumbers,amsmath,amssymb]{revtex4}

\usepackage{graphicx}
\usepackage{dcolumn}
\usepackage{bm}

\begin{document}

\title{The small $x$ behavior of the gluon structure function \\
from total cross sections} 

\author{E.G.S. Luna$^{1,2}$, A.A. Natale$^{1}$ and C.M. Zanetti$^{1}$}
\affiliation{
$^{1}$Instituto de F\'{\i}sica Te\'orica, UNESP, S\~ao Paulo State University, \\
01405-900, S\~ao Paulo, SP, Brazil \\
$^{2}$Instituto de F\'{\i}sica Gleb Wataghin, \\
Universidade Estadual de Campinas,
13083-970, Campinas, SP, Brazil}

\begin{abstract}
Within a QCD-based eikonal model with a dynamical infrared gluon mass scale we discuss how the small $x$ behavior of the gluon
distribution function at moderate $Q^{2}$ is directly related to the rise of total hadronic cross sections. In this model the
rise of total cross sections is driven by gluon-gluon semihard scattering processes, where the behavior of the small $x$ gluon
distribution function exhibits the power law $xg(x,Q^2)= h(Q^2)x^{-\epsilon}$. Assuming that the $Q^{2}$ scale is proportional
to the dynamical gluon mass one, we show that the values of $h(Q^2)$ obtained in this model are compatible with an earlier
result based on a specific nonperturbative Pomeron model. We discuss the implications of this picture for the behavior of
input valence-like gluon distributions at low resolution scales.
\end{abstract}


\maketitle

\section{Introduction}

\label{intro}

The  increase  of  hadronic  total cross  sections  was  theoretically
predicted  many years ago \cite{cheng} and this  prediction has
been accurately verified by experiment \cite{pdg}. Nowadays, one of the
main  theoretical tools to  explain this  behavior is  the diffractive
formalism      underlying     the      so      called     QCD-inspired
models \cite{halzenl,godbole,durand,luna}. In  this formalism the
scattering amplitude  in the c.m.  system is associated  with semihard
processes, i.e. processes with very  small $x$ whose amplitudes can be
calculated  in   perturbative  QCD:  at  high   energies  the  typical
transverse momentum of semihard  processes is relatively large and for
wee partons in hadronic scattering  processes it leads to small values
for  the  strong coupling  constant  $\alpha_{s}$.   At high  energies
semihard processes compete with ``soft'' ones which have traditionally
been regarded as being responsible  for the bulk of the hadronic cross
section.   Hence the  study  of hadronic  scattering  processes is  no
further away from the QCD theory  than, for example, the study of hard
processes  at small  transverse  distances; as  the energy  increases,
semihard processes are expected  to give an increasing and significant
part of  the total hadronic  cross section \cite{gribov,ryskin01}. This
appealing  idea  is  a  feasible  one  since it  is  well  known  that
experimental observations made it clear that at high energies the soft
and the  semihard components  of hadron-hadron scattering  are closely
related, just as, for example, the observation of correlations between
the   average  transverse   momenta  of   hadrons  produced   and  the
multiplicity  density  in   rapidity, \cite{arnison}  and  the  copious
production of gluon jets  with moderate $p_{T}$ (minijet phenomena) in
hadronic collisions \cite{minijet}.

QCD-inspired  models incorporate  soft and  semihard processes  in the
treatment   of  high   energy  hadron-hadron   interactions   using  a
diffraction-scattering  formulation  compatible  with analyticity  and
unitarity  constraints.   At   large  energies  and  fixed  transverse
momentum there  is a  rapid growing number  of semihard gluons  in the
hadron, which  is the main reason  of the rising  total cross section.
Therefore  the rise  of  total hadronic  cross  section is  intimately
connected to the small $x$ behavior of the gluon distribution function
and, within the range of applicability of these models, it can be used
to  investigate  the gluon  distribution  exactly  in  the $x  \ll  1$
semihard region.

In QCD-inspired models a  phenomenological gluon distribution with the
behavior $g(x)\sim h x^{-J}$ is used in the actual calculations, where
$J > 1$.  It could be asked why this approach has not been used before
to determine  the small $x$  behavior of $g(x)$.  One  possible reason
lies in the fact that previous  models were quite dependent on an {\it
ad hoc}  infrared cutoff mass scale  and on the infrared  value of the
strong                        coupling                        constant
($\alpha_0$) \cite{halzenl,godbole,durand}. Therefore,  if $g(x)$
at small  $x$ is given by  the above expression,  the determination of
$h$ would not  be reliable since it appears  multiplied by $\alpha_0$,
which is a fitting parameter in the model.

The  freedom to choose  the values  of the  mass scale  and $\alpha_0$
disappears in one  improvement of the model, where  the arbitrary mass
scale  is  changed  by  a  dynamical  one \cite{luna}.  In  this  model
(henceforth referred  to as DGM model)  the onset of  the dominance of
semihard gluons  in the interaction of high-energy  hadrons is managed
by the dynamical gluon mass $m_g$ \cite{cornwall}
(intrinsically    related     to    an    infrared     finite    gluon
propagator, \cite{ans}) whose existence is strongly supported by recent
QCD  lattice simulations \cite{lqcd} as  well as  by
phenomenological
results \cite{luna,cornwall,aguilar,halzen,luna8,luna3};
for  recent  reviews  on  the  phenomenology  of  massive  gluons  see
Ref. \cite{luna10} and  references therein.  In the
DGM model  the infrared  coupling constant is  also a function  of the
dynamical  gluon  mass \cite{cornwall}. Therefore
we have a precise physical  meaning for this infrared coupling as well
as for the  quoted dynamical mass scale, which  is a natural regulator
of the  infrared divergences associated with  the semihard gluon-gluon
subprocess  cross sections.   These  properties render  the DGM  model
instrumental for  a reliable determination of  the QCD-predicted small
$x$ distribution  $g(x,Q^2)= h(Q^2) x^{-J}$ at  moderate $Q^2$, namely
the determination  of the $h(Q^2)$ factor,  where the $Q$  scale is on
the order of  the dynamical gluon mass $m_{g}$.   This dynamical scale
represents the onset of semihard contributions to hadronic diffractive
scattering and  hence provides the  necessary scale in order  to apply
the diffractive formalism.

The  gluon distribution function  is usually  determined in  the large
$Q^2$  region  and  its  small  $x$  limit in  this  region  has  been
extensively discussed  in recent times. Therefore the  results that we
will  discuss  in  this  work  can  be  considered  phenomenologically
important if we  want to understand the match  between these different
kinematical regions.  Moreover, the  gluon distribution at a low value
of $Q^2=Q_{0}^2$  has to be taken  into account in  order to calculate
the    distribution   at    higher   $Q^2$    via    DGLAP   evolution
equations \cite{dglap}. We  show that the  origin of
such low  $Q_{0}^2$ initial distribution  can be naturally  related to
the phenomenon of dynamical mass generation of gluons.

The paper  is organized as follows:  in the next  section we introduce
the DGM model and show  how the semihard gluon-gluon scattering can be
properly  eikonalized and  linked to  $pp$ and  $\bar{p}p$ diffractive
scattering.  Our results  for $h(Q^2)$  at moderate  $Q^2$  region are
presented  in the  Sec. III,  where we  also address  the  question of
extrapolating  the  gluon distribution  function  beyond the  infrared
$Q^2$ range.  In Sec. IV we draw our conclusions.

\section{Gluon distribution and rising cross sections}

In order to set a natural  moderate scale $Q^2$ and produce a reliable
determination  of  the  factor  $h(Q^2)$,  we explore  the  small  $x$
behavior of $g(x,Q^2)= h(Q^2)  x^{-J}$ in one eikonalized QCD-inspired
model  improved with  a  dynamical gluon  mass \cite{luna}. This  model
(previously  referred   to  as   DGM  model)  provides   a  consistent
calculation  of total  cross sections  consonant with  analyticity and
unitarity constraints,  and has  already been applied  successfully to
describe $pp$  and $\bar{p}p$ diffractive  scattering data \cite{luna},
as well  as data  on $\gamma p$  photoproduction and  hadronic $\gamma
\gamma$   total    cross   sections \cite{luna3}. In    the   eikonal
representation the total cross section is given by
\begin{eqnarray}
\sigma_{tot}(s)   =  4\pi   \int_{_{0}}^{^{\infty}}   \!\!  b\,   db\,
[1-e^{-\chi_{_{I}}(b,s)}\cos \chi_{_{R}}(b,s)],
\label{eq21}
\end{eqnarray}
where $s$ is the square of the total center-of-mass energy, $b$ is the
impact  parameter,  and $\chi(b,s)=\chi_{_{R}}(b,s)+i\chi_{_{I}}(b,s)$
is a  complex eikonal  function. In  the DGM model  we write  the even
eikonal  as  the  sum  of gluon-gluon,  quark-gluon,  and  quark-quark
contributions:
\begin{eqnarray}
\chi^{+}(b,s) &=&  \chi_{qq} (b,s) +\chi_{qg} (b,s)  + \chi_{gg} (b,s)
\nonumber  \\  &=&  i[\sigma_{qq}(s)  W(b;\mu_{qq})  +  \sigma_{qg}(s)
W(b;\mu_{qg}) \nonumber \\ & & + \sigma_{gg}(s) W(b;\mu_{gg})] ,
\label{eq22}
\end{eqnarray}
where $\chi_{pp}^{\bar{p}p}(b,s) = \chi^{+} (b,s) \pm \chi^{-} (b,s)$.
Here $W(b;\mu)$ is the overlap  function in the impact parameter space
and $\sigma_{ij}(s)$  are the elementary subprocess  cross sections of
colliding  quarks  and gluons  ($i,j=q,g$).  The  overlap function  is
associated with the Fourier transform of a dipole form factor,
\begin{eqnarray}
W(b;\mu) = \frac{\mu^2}{96\pi}\, (\mu b)^3 \, K_{3}(\mu b),
\label{eq23}
\end{eqnarray}
where $K_{3}(x)$ is  the modified Bessel function of  second kind. The
odd eikonal $\chi^{-}(b,s)$, that  accounts for the difference between
$pp$ and $\bar{p}p$ channels, is parametrized as
\begin{eqnarray}
\chi^{-} (b,s)  = C^{-}\, \Sigma \,  \frac{m_{g}}{\sqrt{s}} \, e^{i\pi
/4}\, W(b;\mu^{-}),
\label{eq24}
\end{eqnarray}
where $m_{g}$  is the  dynamical gluon mass  (which will  be discussed
afterward) and the parameters  $C^{-}$ and $\mu^{-}$ are constants to
be fitted. The factor $\Sigma$ is defined as
\begin{eqnarray}
\Sigma = \frac{9\pi \bar{\alpha}_{s}^{2}(0)}{m_{g}^{2}},
\label{eq25}
\end{eqnarray}
with  the dynamical  coupling constant  $\bar{\alpha}_{s}$ set  at its
frozen infrared  value.  The  eikonal functions $\chi_{qq}  (b,s)$ and
$\chi_{qg} (b,s)$, needed to describe the low-energy forward data, are
parametrized  with terms  dictated  by the  Regge phenomenology.   The
formal expressions  of these quantities as well  as details concerning
the  analyticity properties  of the  model amplitudes  can be  seen in
Ref. \cite{luna}.

In the DGM  model the main contribution to  the asymptotic behavior of
hadron-hadron  total cross  sections comes  from  semihard gluon-gluon
collisions.   The gluon  eikonal term  $\chi_{gg}(b,s)$ is  written as
$\chi_{gg}(b,s)\equiv \sigma_{gg}(s)W(b; \mu_{gg})$, where
\begin{eqnarray}
\sigma_{gg}(s) = \int_{4m_{g}^{2}/s}^{1} d\tau \,F_{gg}(\tau )\,
\hat{\sigma}^{D\!PT} (\hat{s}) .
\label{eq28}
\end{eqnarray}
Here  $F_{gg}(\tau )$ is  the convoluted  structure function  for pair
gluon-gluon ($\tau = x_{1}x_{2}$), and $\hat{\sigma}^{D\!PT}(\hat{s})$
is  the  complete  cross   section  for  the  subprocess  $gg\to  gg$,
calculated     through    the     dynamical     perturbation    theory
(DPT) \cite{pagels}, where the  effective gluon propagator and coupling
constant enters into the  calculation. These effective quantities were
obtained as  solutions of the Schwinger-Dyson equations  (SDE) for the
gluon   propagator   and   triple   gluon   vertex   via   the   pinch
technique \cite{cornwall}, and  have been used in
many phenomenological calculations that are sensible to their infrared
finite  behavior \cite{aguilar,halzen}. In this  approach
the  nonperturbative  running   coupling  $\bar{\alpha}_{s}$  and  the
functional expression of the  dynamical gluon mass $M_{g}^2 (q^2)$ are
given by
\begin{eqnarray} 
\bar{\alpha}_{s} (q^2)= \frac{4\pi}{\beta_0 \ln\left[
(q^2 + 4M_g^2(q^2) )/\Lambda^2 \right]}, 
\label{eq11}
\end{eqnarray}
\begin{eqnarray}
M^2_g(q^2) =m_g^2 \left[\frac{ \ln
\left(\frac{q^2+4{m_g}^2}{\Lambda ^2}\right) } {
\ln\left(\frac{4{m_g}^2}{\Lambda ^2}\right) }\right]^{- 12/11} ,
\label{mdyna} 
\end{eqnarray}
respectively, where $\Lambda$($\equiv\Lambda_{QCD}$)  is the QCD scale
parameter, $q$ is the  four-momentum transfer between the incoming and
outgoing  parton, and  $\beta_0 =  11- \frac{2}{3}n_f$  ($n_f$  is the
number of flavors).  The dynamical  gluon mass $m_{g}$ has to be found
phenomenologically, and a  typical value (for $\Lambda =  300$ MeV) is
$m_g         \approx          500         \pm         200$         MeV
\cite{cornwall,aguilar,halzen,luna8,luna3}. The
expression  of $M_g^2  (q^2)$ comes  out as  a solution  of  the gluon
propagator SDE and  $m_g$ is interpreted as the  gluon dynamical mass,
being related to the value of the propagator at $q^2=0$. This mass has
a dynamical origin, which, in principle, should be calculated in terms
of the scale $\Lambda$,  and in this work it will be  used as an input
parameter.   As discussed  in detail  in the  Ref. \cite{luna}, the
total                           cross                          section
$\hat{\sigma}^{D\!PT}(\hat{s})=\int_{\hat{t}_{min}}^{\hat{t}_{max}}
(d\hat{\sigma}/d\hat{t}\,   )   \,   d\hat{t}$  for   the   subprocess
$gg\rightarrow gg$  is obtained  by integrating over  $4m_g^2 -\hat{s}
\le  \hat{t} \le  0$.  In  setting  these kinematical  limits we  have
neglected   the    momentum   behavior   in    Eqs.~(\ref{eq11})   and
(\ref{mdyna}),  since the  calculation of  the hadronic  cross section
does not depend  strongly on the specific form  of $M_{g}(q^{2})$, but
more  on  its  infrared  value   (i.e.   the  value  of  $m_{g}$).   A
straightforward calculation yields
\begin{eqnarray}
\hat{\sigma}^{D\!PT}(\hat{s})              =              \frac{3\pi
  \bar{\alpha}_{s}^{2}}{\hat{s}}   \left[  \frac{12\hat{s}^{4}   -  55
  m_{g}^{2}  \hat{s}^{3} +  12  m_{g}^{4} \hat{s}^{2}  + 66  m_{g}^{6}
  \hat{s}   -   8    m_{g}^{8}}{4   m_{g}^{2}   \hat{s}   [\hat{s}   -
  m_{g}^{2}]^{2}}  -     3     \ln     \left(    \frac{\hat{s}     -
  3m_{g}^{2}}{m_{g}^{2}}\right) \right] .
\label{h6}
\end{eqnarray}

In  the calculation  of  the structure  function $F_{gg}(\tau  )\equiv
[g\otimes g](\tau )$ we  adopt a QCD-motivated distribution functional
form:
\begin{eqnarray}
g(x,Q^2) = h(Q^2) \, \frac{(1-x)^5}{x^{J}},
\label{dist33}
\end{eqnarray}
where the  quantity $J\equiv 1+\epsilon  > 1$ controls  the asymptotic
behavior of  $\sigma_{tot}(s)$. The factor $(1-x)^5$  is introduced in
accord with the spectator  counting rules at high $x$ \cite{brodsky01}.
In  the  Regge language  the  quantity $J$  is  the  intercept of  the
Pomeron.  As  we go  up in energy  ($s \gg \Lambda^2$  and transferred
momenta $Q^2 \approx m_{g}^2$) the  proton will be filled up with more
and more semihard gluons, and we will be probing the small $x$ region.
Therefore we see that the behavior of the total hadronic cross section
at higher  energies is driven  by only four parameters:  $m_{g}$, $J$,
$h$ and $\mu_{gg}$  (see Eqs.~(\ref{eq21}), (\ref{eq22}), (\ref{eq23})
and (\ref{dist33})).   The first two are input  parameters whereas $h$
and $\mu_{gg}$ are fitting ones.  As we will see in detail in the next
section, $h$ and  $\mu_{gg}$ are determined carrying out  a global fit
to high-energy $pp$ and $p\bar{p}$ scattering data, namely total cross
section $\sigma_{tot}$ and $\rho$ parameter data.

In  the DGM  model  the  connexion between  the  dynamical gluon  mass
$m_{g}^2$ and  the scale $Q^2\equiv  -q^2$ arises as follows:  the QCD
improved parton  model is not applicable  only in the  region in which
the transverse momenta $p_{T}$ of the colliding hadrons is of the same
order as their center-of-mass energy,  but also in the region in which
$\sqrt{s}\gg        p_{T}$        provided       that        $p_{T}\gg
\Lambda_{QCD}$ \cite{gribov,ryskin01}. Hence   it   is  possible   to
investigate the interface between  high $p_{T}$ region and Regge limit
one by means  of perturbative methods.  In the  DGM model we introduce
the  physical energy  threshold $s=  4m_{g}^{2}$ for  the  final state
gluons (see expression (\ref{eq28})), assuming that these are screened
gluons,   in    a   procedure   similar   to    the   calculation   of
Ref. \cite{cornwall2}.    In  this   way,  adopting   the  relation
$p_{T}^2=Q^2$, we establish the expected particle production threshold
$Q=2m_{g}$. Note  that the gluon  distribution function (\ref{dist33})
reduces  to  $g(x,Q^2)=0$  in  the  limit $x\to  1$,  as  expected  by
dimensional  counting  rules, and,  more  importantly, reproduces  the
small $x$ QCD prediction
\begin{eqnarray}
g(x,Q^2) = h(Q^2) \, {x^{-J}}.
\label{dist45}
\end{eqnarray}
In this context, given the high-energy nature of the data to be fitted
in  the  following  analysis,  we  are specially  concerned  with  the
determination  of $g(x,Q)$  in its  small $x$  region  and transferred
momenta of the same order as  the dynamical gluon masses.  As we shall
discuss  in   the  next  section   the  cross  section   described  by
Eq.~(\ref{eq28}) is totally  dominated by the small $x$  region and by
processes   at  $Q^2   \approx   4m_g^2$.  However   $m_g  \approx   2
\Lambda_{QCD}$           ,          as           discussed          in
Ref. \cite{cornwall} (although the uncertainty
in  this value is  of $O(\Lambda_{QCD})$),  and we  are in  a moderate
region of momenta, and  above the nonperturbative region determined by
the QCD scale ($\Lambda_{QCD}$).

\section{The gluon distribution at moderate resolution $Q^2$ scales}

In  the determination  of  the  small scale  dependence  of the  gluon
distribution  we explore  the small  $x$ behavior  of $g(x,Q)$  in one
eikonalized QCD-based  model improved with a dynamical  gluon mass. We
carry out  global fits to  high-energy $pp$ and  $\bar{p}p$ scattering
data above  $\sqrt{s}=10$ GeV  in order to  determine phenomenological
values  for the  factor  $h$ as  a  function of  dynamical gluon  mass
$m_{g}$. Note  that at this $\sqrt{s}$ value  the dynamics responsible
for  the cross  section  increase with  energy  is already  effective.
These data  sets include the total cross  section ($\sigma_{tot}$) and
the  ratio of the  real to  imaginary part  of the  forward scattering
amplitude  ($\rho$ parameter).   We  use the  data  sets compiled  and
analyzed by the Particle  Data Group \cite{pdg}, with the statistic and
systematic  errors added  in  quadrature. In  all  the fits  we use  a
$\chi^{2}$     fitting     procedure,     adopting     an     interval
$\chi^{2}-\chi^{2}_{min}$ corresponding, in the case of normal errors,
to the  projection of the  $\chi^{2}$ hypersurface containing  90\% of
probability.  In  the case  of 8 fitting  parameters (DGM  model) this
corresponds to the interval $\chi^{2}-\chi^{2}_{min}=13.36$.

As discussed in  the previous section, the behavior  of the total $pp$
and $p\bar{p}$  cross sections at higher energies  is driven specially
by  four parameters:  $m_{g}$,  $J$, $h$  and  $\mu_{gg}$.  The  input
values  of  the  $m_{g}$ have  been  chosen  to  lie in  the  interval
$[350,750]$ MeV, as suggested  by the value $m_{g}=400^{+350}_ {-100}$
MeV  obtained  in a  previous  analysis  of  the $pp$  and  $\bar{p}p$
channels through the DGM  model \cite{luna}. This input dynamical gluon
mass  range is  also  supported by  recent  studies via  the model  on
survival    probability    of    rapidity    gaps    in    diffractive
scattering \cite{luna8} as  well as  on the $\gamma  p$ photoproduction
and the hadronic $\gamma  \gamma$ total cross sections \cite{luna3}. We
stress that if we had used  $m_g$ as another parameter to be fitted we
would  reproduce  the  results  of Fig.(1)  of  Ref. \cite{luna},
indicating  that the  best gluon  mass values  would be  in  the range
$[300,750]$ MeV.

A  consistent  input  value  of  $J$  can  be  obtained  from  fitting
scattering data via Regge phenomenology:  in the limit of large enough
$s$ the gluon-gluon cross  section $\sigma_{gg}(s)$ obtained using the
gluon  distribution  (\ref{dist45}) behaves  as  a  Pomeron power  law
$s^{J-1}$:
\begin{eqnarray}
\lim_{s\to  \infty}  \int^{1}_{4m_{g}^{2}/s}  d\tau \,  F_{gg}(\tau)\,
\hat{\sigma}^{D\!PT}   (\hat{s})   \sim  \left(   \frac{s}{4m_{g}^{2}}
\right)^{\epsilon} ,
\label{gtso2}
\end{eqnarray}
i.e.   the quantity  $\epsilon$  controls the  asymptotic behavior  of
$\sigma_{gg}(s)$.    We   see   that   the  asymptotic   behavior   of
$\sigma_{gg}(s)$ in the DGM model  is similar to the one expected from
the  Regge phenomenology,  where the  quantity $J$  is related  to the
intercept    of   the    Pomeron,   with    $J\equiv    1+\epsilon   >
1$ \cite{pomeron}. From  fitting scattering  data
through an  extended Regge model,  the value of the  Pomeron intercept
$J$ imposed by the accelerator data currently available is $J=1.088\pm
0.011$ \cite{luna2}. This value is compatible
with the recent result $J=1.085\pm 0.006$, obtained through the use of
non-linear trajectories for  the meson resonances \cite{luna4}. Also, a
Regge fit of the proton structure function data indicates the presence
of  a   second  Pomeron,  with   the  hard  intercept   $1+\epsilon  =
1.435$ \cite{landshoff5}. This  last value of  $\epsilon$ is compatible
with the one expected for the exponent $\lambda$ from the perturbative
BFKL         approach \cite{bfkl},         namely        $\lambda
=\frac{12}{\pi}\alpha_{s}\ln 2 \approx 0.5$.  However, there is a lack
of  theoretical precision about  the precise  value of  $\lambda$ once
their  value is  dependent  upon an  infrared  cut-off $k_{0}^{2}$  at
transverse momenta of gluons \cite{kwi,forshaw2}. Moreover,
it  is also  possible  that  the phenomenon  of  dynamical gluon  mass
generation push the  perturbative $\lambda$ value to a  smaller one as
was  shown years ago  by Ross \cite{ross1}. More importantly,  in what
follows  we will show  that the  use of  a hard  value to  the Pomeron
intercept gives a bad fit to $\sigma_{tot}$ and $\rho$ parameter data.
It  shows that  in  fact  the increase  of  hadron-hadron total  cross
sections is governed  by soft values of $J$,  as already expected from
Regge and analytical models \cite{landshoff2,luna5}.

The   input   values   for   $J$   have  been   extracted   from   the
Ref. \cite{luna2}, where an extended Regge parametrization was used
in   order  to  establish   extrema  bounds   for  the   soft  Pomeron
intercept. In  that analysis  the intercept of  the soft  Pomeron were
determined  exploring  the  systematic  uncertainty  coming  from  the
discrepancy  concerning the  results for  $\sigma_{tot}^{\bar{p}p}$ at
$\sqrt{s}=1.8$  TeV reported  by the  CDF  Collaboration \cite{CDF} and
those  reported  by the  E710  \cite{E710}  and  the E811  \cite{E811}
Collaborations; the  values for  the lower and  upper bounds are  $J =
1.085$ and $J = 1.095$, respectively, and represent the bounds for the
soft Pomeron intercept imposed by  the accelerator and cosmic ray data
currently  available \cite{luna2}. We shall  restrict ourselves  to the
above range  of $J$  values.  Values outside  this window  provide bad
fits to the experimental data.

\begin{table*}
\caption{Values of the input parameters and fitting ones of the DGM model resulting from the global fit to the $pp$ and
$\bar{p}p$ scattering data.}
\begin{ruledtabular}
\begin{tabular}{cccc}
$m_{g}$ [MeV] & $\mu_{gg}$ [GeV] & $h$ & $\chi^2/DOF$\\
\hline
\multicolumn{4}{c}{$J=1+\epsilon=1.085$} \\
300 & 0.741$\pm$0.059 & 0.162$\pm$0.009 & 1.112 \\
350 & 0.782$\pm$0.064 & 0.226$\pm$0.013 & 1.111 \\
400 & 0.811$\pm$0.065 & 0.292$\pm$0.018 & 1.101 \\
450 & 0.857$\pm$0.073 & 0.357$\pm$0.021 & 1.114 \\
500 & 0.781$\pm$0.061 & 0.423$\pm$0.028 & 1.115 \\
550 & 0.878$\pm$0.072 & 0.525$\pm$0.037 & 1.107 \\
600 & 0.920$\pm$0.074 & 0.634$\pm$0.041 & 1.108 \\
650 & 0.909$\pm$0.071 & 0.708$\pm$0.052 & 1.107 \\
700 & 0.868$\pm$0.078 & 0.841$\pm$0.054 & 1.124 \\
750 & 0.956$\pm$0.078 & 0.993$\pm$0.065 & 1.124 \\
\hline
\multicolumn{4}{c}{$J=1+\epsilon=1.435$} \\
400 & 0.679$\pm$0.139 & 0.017$\pm$0.001 & 1.950 \\
\end{tabular}
\end{ruledtabular}
\end{table*}

The results  of our global  fits to the  data set discussed  above are
depicted in Fig. 1. We note a fast increase in the $h$ values with the
dynamical gluon mass.  This behavior  does not depends on the specific
value of the soft intercept,  since the results are totally compatible
for  the two  intercept bounds,  namely $J=1.085$  and  $J=1.095$. The
$\chi^2/DOF$  values obtained in  the global  fits with  $J=1.085$ are
relatively low,  as show in Table  I (the $\chi^2/DOF$  values for the
input  $J=1.095$ are  similar).   These results  (for  148 degrees  of
freedom) indicate the excellence of the fits and show that it is quite
unlikely that the soft Pomeron does not dominate the behavior of total
cross  sections   at  least   in  the  energy   region  that   we  are
considering. In other  words, our fits do not  support a hard Pomeron,
as indicated in the Table I: the use of the hard input value $J=1.435$
results  in a  significantly worse  fit ($\chi^2/DOF=1.95$)  and shows
that in this case the DGM model does not accommodate the data set used
in  the  fitting  procedure.   Moreover, the  input  choice  $J=1.435$
results   in  a   very  low   value   for  the   factor  $h$,   namely
$h=0.017\pm0.001$. As we  will see in the next  section, this value is
approximately 1/10 of the expected one \cite{cudell}.

The small $x$ behavior of $xg(x,Q^2)=h(Q^2)x^{-\epsilon}$ at $Q^2=1$
GeV$^2$, taking into account our $h$ result for $m_{g}=500$ MeV as
indicated in Table I, is given by
\begin{eqnarray}
xg(x) = (0.423\pm 0.028)x^{-(0.085\pm 0.05)},
\label{disour}
\end{eqnarray}
where we have adopted the intercept bound $1+\epsilon =1.085$ which is
consistent  with the recent  value of  $J$ obtained  from spectroscopy
data analysis \cite{luna4}. This behavior can be compared with the MRST
result for the LO parametrization of $g(x)$ at $Q^2_0 =1$ GeV$^2$
\noindent at very small $x$ \cite{mrst}:
\begin{eqnarray}
xg^{MRST}(x) \approx 3.08x^{0.10};
\label{mr}
\end{eqnarray}
at   $x'=10^{-5}$  (the  kinematical  limit   of  the  MRST
distributions)            our            distribution            gives
$x'g(x')=1.125^{+0.146}_{-0.133}$,   whereas   $x'g^{MRST}(x')=0.974$.
This numerical  agreement reinforces  the assertion that  the hadronic
scattering is driven  mainly by soft Pomerons at  scales $Q^2 \lesssim
1$ GeV$^2$. This is  not the case for higher scales, as  we can see in
Figure  2,   where  we  compare   our  gluon  distribution   with  the
MRST \cite{mrst} and CTEQ6L1,  CTEQ6L \cite{cteq6} leading-order
gluon distributions at $Q^2=1.69$  GeV$^2$ and $Q^2=4$ GeV$^2$.  These
parton  distributions are  evolved with  LO splitting  functions using
DGLAP equations from the initial scales $Q_{0}^2=1$ GeV$^2$ (MRST) and
$Q_{0}^2=1.69$  GeV$^2$  (CTEQ6), and  are  obtained using  tree-level
formulas for the hard cross sections. The CTEQ6 (MRST) parton sets are
valid in the kinematical intervals $10^{-6} \le x \le 1$ ($10^{-5} \le
x \le  1$) and 1.3 GeV  $\le Q \le 10^4$  GeV (1 GeV $\le  Q \le 10^7$
GeV).   The  CTEQ6L   distribution  uses  the  same  $\alpha_{s}(Q^2)$
coupling as  the standard next-to-leading  order (NLO) CTEQ  fits with
$\alpha_{s}(M_{Z}^2)$=0.118  whereas CTEQ6L1 one  uses the  LO formula
for  $\alpha_{s}(Q^2)$ with $\Lambda_{QCD}^{4flavor}=0.215$  GeV (with
corresponds to $\alpha_{s}(M_{Z}^2)=0.130$). The LO MRST distributions
uses  a value  of $\Lambda^{4flavor}_{QCD}=220$  MeV  corresponding to
$\alpha_{s}(M_{Z}^2)\approx 0.130$. In the  DGM model we adopt the QCD
parameter  value  $\Lambda(n_{f}=4)=300$  MeV,  which  corresponds  to
$\bar{\alpha}_{s}(M_{Z}^2)\approx  0.130$.   The  results depicted  in
Figure 2 indicate that the small-$x$ gluons at higher $Q^2$ scales are
mainly  generated radiatively: at  the value  $Q^2=4$ GeV$^2$  the DGM
gluon distribution  and the CTEQ/MRST  ones are discrepant  at greater
than a factor  of 5 at $x=10^{-5}$, although the  DGM and MRST results
are compatible  at $Q^2=1$ GeV$^2$ scale. The  DGM gluon distributions
at higher scales  can be calculated by fitting the  $h$ results to the
functional for $h(Q^2)=A+BQ^2$,  as shown in Figure 3,  where the best
fit   gives  $A=0.121\times   10^{-1}$  and   $B=0.170\times  10^{-5}$
MeV$^{-2}$. It  is clear that the  functional form of  $h(Q^2)$ is not
known and we stress that the choice  we made above is just a guess, as
well as it  should be kept in  mind that the $Q^2$ values  that we are
using in this function are of the order of $4m_g^2$.

It is worth  mentioning that mass values $m_{g}  \lesssim 500$ MeV are
favored in other calculations of strongly interacting processes, as we
will  discuss later.   In this  way,  we consider  the more  stringent
moderate-$Q^2$  interval 0.36  GeV$^2$  $\le Q^2  \lesssim 1$  GeV$^2$
(corresponding to the gluon mass  interval 300 MeV $\le m_{g} \lesssim
$ 500 MeV) as a more reliable one for the DGM gluon distribution. This
new  threshold interval  for the  production of  dynamically generated
gluons is  corroborated by our  results at $Q^2 \gtrsim  1.69$ GeV$^2$
($m_{g}\gtrsim 650$ MeV) since  our gluon distribution is smaller than
the LO CTEQ6 and  MRST ones. It is a clear indication  that in the DGM
model threshold  values $m_{g}\gtrsim 500$ MeV  inhibit the production
of nonperturbative gluons.

At even smaller scales, for instance $Q^2 \lesssim $ 0.36 GeV$^2$, the
$x$ behavior of  the gluon distribution is due to  the dynamics of the
bound  state proton  and  is an  unsolved  problem of  nonperturbative
QCD. Hence we guess a  behavior for $g(x,Q^2)$ that provides a picture
more  consistent with  the  expected behavior  of the  nonperturbative
dynamics            of            the           bound            state
nucleon \cite{magnin,ingelman}: we  assume the frozen form
$g(x,\eta^2)$ for momenta  $Q\le \eta = 300$ MeV. In  this way, we can
write  down  a  phenomenologically  useful  expression  to  the  gluon
distribution function in the range $Q \lesssim $ 1 GeV:
\begin{eqnarray}
g(x,Q^2) = h(\eta^2) \, \frac{(1-x)^5}{x^{J}} \, \theta (\eta^2 - Q^2)
+ h(Q^2) \, \frac{(1-x)^5}{x^{J}} \, \theta (Q^2 - \eta^2 ) .
\label{dist678}
\end{eqnarray}

This phenomenological  gluon distribution takes into  account the fact
that at small $Q^2$ scales ($Q \lesssim 300$ MeV) the usual methods of
analysis  of parton  distribution based  on perturbative  QCD  are not
applicable. It is clear that  low-$Q^2$ photons do not resolve partons
in the proton and the deep inelastic scattering (DIS) formalism cannot
be  applied  straightforwardly,  and  probably  a  generalized  vector
dominance model, as  discussed by Alwall and Ingelman \cite{ingelman2},
is more appropriate at low $Q^2$. Generally speaking we could say that
we  should expect a  smaller gluon  distribution function  at moderate
$Q^2$ than the one  obtained through a strictly perturbative technique
(the  CTEQ approach,  for  instance), because  of  the propagator  and
coupling constant softening  in the infrared which will  appear in the
cross section calculation \cite{ans,aguilar}.

\section{Conclusions}

From  the diffractive  formalism underlying  QCD-based models  we have
showed how the rise of  total hadronic cross sections can be naturally
connected  to  the  small  $x$  behavior of  the  QCD-predicted  gluon
distribution function $g(x,Q)$ at moderate $Q^2$.  We have performed a
precise  determination of  the  $h(Q^2)$ factor  that  appears in  the
QCD-predicted gluon  distribution in the framework of  the DGM eikonal
model.   This   QCD-based  model   is  instrumental  for   a  reliable
determination of the small $x$  behavior of $g(x,Q)$ at moderate $Q^2$
since  diffractive processes  involve  both semihard  and soft  scales
resulting  in   a  complicated  interplay   between  perturbative  and
nonperturbative  effects. More  specifically, we  have  determined the
behavior of  the gluon distribution  $g(x,Q^2)= h(Q^2) x^{-J}$  in the
moderate $Q$  interval $Q \lesssim  1$ GeV. The connexion  between the
resolution scale  $Q$ and  the gluon dynamical  mass $m_{g}$  has been
established by the particle production threshold $Q=2m_{g}$.

Our results show that at momenta $Q^2 \gtrsim 1$ GeV$^2$ the small-$x$
gluons are mainly generated  radiatively. The coefficient $h$ is $Q^2$
dependent as  discussed at the end  of the last  section.  We verified
that $J$ values like  $J=1+\epsilon\approx 1.08$ are preferred instead
of  the large (hard  perturbative Pomeron)  value $J=1+\epsilon\approx
1.43$. However the rapid increase of $h(Q^2)$ with the momentum is not
enough to  produce a number of  nonperturbative gluons as  high as the
perturbative ones. For this is  probably necessary a rapid increase of
the $J$  values with the momentum in  such a way that  soft values are
preferred at moderate  $Q^2$ and hard ones would  be necessary only at
high  $Q^2$. We  call attention  to the  fact that  this  statement is
corroborated  by a  MRST  group analysis  of  parton distributions  of
proton \cite{mrst2}. From fitting  the  sea quark  (S)  and gluon  (G)
distributions of the default  MRST partons to the forms $xf_{i}(x,Q^2)
= A(Q) x^{-\lambda_{i}(Q^2)}$ as $x\to  0$, $i=S,G$, they have look at
how the exponents $\lambda_{S}$  and $\lambda_{G}$ vary with $Q^2$. It
was observed  that as $Q^2$  increases from the input  scale $Q_{0}=1$
GeV$^2$ the valence-like character of the gluon rapidly disappears due
to evolution being  driven by the much steeper  sea. For higher values
of  $Q^2$  the  gluon  exponent $\lambda_{G}$  increases  rapidly  and
becomes higher  than the sea  quark exponent $\lambda_{S}$,  since the
gluon  drives the  sea quark  via the  $g\to \bar{q}q$  transition. By
taking  only positive  values  for $\lambda_{G}$  we  observe that  it
starts at a value where the gluon distribution is almost flat, that is
$\lambda_{G} \approx  0$, and  by $Q^2 \approx  4$ GeV$^2$ it  has the
value  $\lambda_{G}=0.2$.   The   hard  value  $\lambda_{G}=0.435$  is
achieved  at $Q^2  \approx 700$  GeV$^2$.  Therefore  our  results are
consistent with  the conventional  view that the  steeply-rising gluon
component is  absent at  moderate $Q^2$ and  as $Q^2$ increases  it is
generated through pQCD evolution.

The  small $x$ behavior  of the  gluon distribution  (specifically the
value  of  $h$) was  performed  in  a  different approach  by  Cudell,
Donnachie and  Landshoff (CDL) \cite{cudell}, within  the Landshoff and
Nachtmann Pomeron model  (LN) \cite{ln,sho,shoa}. In this model
the small $x$ gluon distribution function is given by
\begin{eqnarray}
xg(x) \equiv h = \frac{4\alpha_s}{3\pi} \int dk^2 k^2 D^2 (-k^2), 
\label{eq15}
\end{eqnarray}
where  the  gluon propagator  ($D$)  is  the  effective one  with  the
perturbative   part  subtracted  from   the  integral.    Notice  that
Eq.~(\ref{eq15})  does not  contain  the $x^{-\epsilon}$  contribution
which has  to be put by hand  in the LN model.   This calculation with
Cornwall's   propagator,  where  we   can  substitute   $\alpha_s$  by
$\bar{\alpha_s}$,  gives $h  \approx  0.15$: this  value is  basically
equal to the value obtained in CDL approach \cite{cudell}. A comparison
of our results with the CDL one shows that our values of $h$ are quite
reasonable.  More explicitly,  shows that  the  CDL result  of $h$  is
compatible with  the DGM  one at  mass scales on  the order  of $m_{g}
\approx 300$ MeV  (see Table I). Dynamical gluon  masses of that order
where obtained in Ref. \cite{luna8} in order to obtain the survival
probability $\langle |S|^{2} \rangle$  of rapidity gaps in diffractive
scattering  from  production  of  Higgs  boson via  $WW\to  H$  fusion
processes.  These survival factors  are compatible with recent results
of  $\langle |S|^{2}  \rangle$ obtained  by Khoze,  Martin  and Ryskin
through   a  two-channel  eikonal   model  which   embodies  pion-loop
insertions  in the Pomeron  trajectory, diffractive  dissociation, and
rescattering effects \cite{khoze01}. In addition, in  the analysis via
DGM model of  the $\gamma p$ photoproduction and  the hadronic $\gamma
\gamma$  total  cross  sections,   both  derived  from  the  $pp$  and
$p\bar{p}$  scattering  amplitudes  assuming vector  meson  dominance,
there is  room for smaller values  of $m_{g}$ beside  the adopted mass
scale $m_{g}=400$  GeV \cite{luna3}. All  these results suggest  a most
restrictive  mass  range  than  the  one  we  have  considered  up  to
now. Given the behavior of  $g(x,Q^2)$ obtained through the DGM model,
we can  be confident in the mass  range $m_{g} \lesssim 500$  MeV as a
more reliable one.  This corresponds  to the $Q^2$ scale interval $Q^2
\lesssim 1$  GeV$^2$. It is clear  that the dynamical gluon  mass is a
concept that  has not been  made totally compatible  with perturbative
QCD, this will  probably demand time and a  great effort, although the
introduction  of   this  concept   is  clearly  consistent   with  the
experimental data and  indicate a more appealing view  to the behavior
of $g(x)$ at small $x$ and moderate $Q^2$.

In  summary,  we  think that  the  DGM  model  is  an useful  tool  in
determining  the  behavior  of  the  gluon  distribution  function  at
moderate $Q^2$ by  merging in a natural way the  soft and the semihard
components of hadronic scattering processes. We emphasize that it is a
nontrivial result that our gluon distribution behavior at small $x$ is
on the order of the MRST one at $Q^2=1$ GeV$^2$. This is achieved with
only  three parameters,  namely  $m_{g}$, $J$  and  $h$.  The  natural
relation between the  momentum scale $Q$ and the  dynamical gluon mass
$m_{g}$ argues the dynamical mass generation of partons to be the main
mechanism  behind  the nonperturbative  dynamics  of the  valence-like
parton distributions present at low resolution scales.

\begin{acknowledgments}
This   research   was   supported   by  the   Conselho   Nacional   de
Desenvolvimento Cient\'{\i}fico  e Tecnol\'ogico-CNPq (EGSL  , AAN and
CMZ).
\end{acknowledgments}

\newpage

\begin{figure}
\resizebox{0.97 \textwidth}{!}{%
\includegraphics{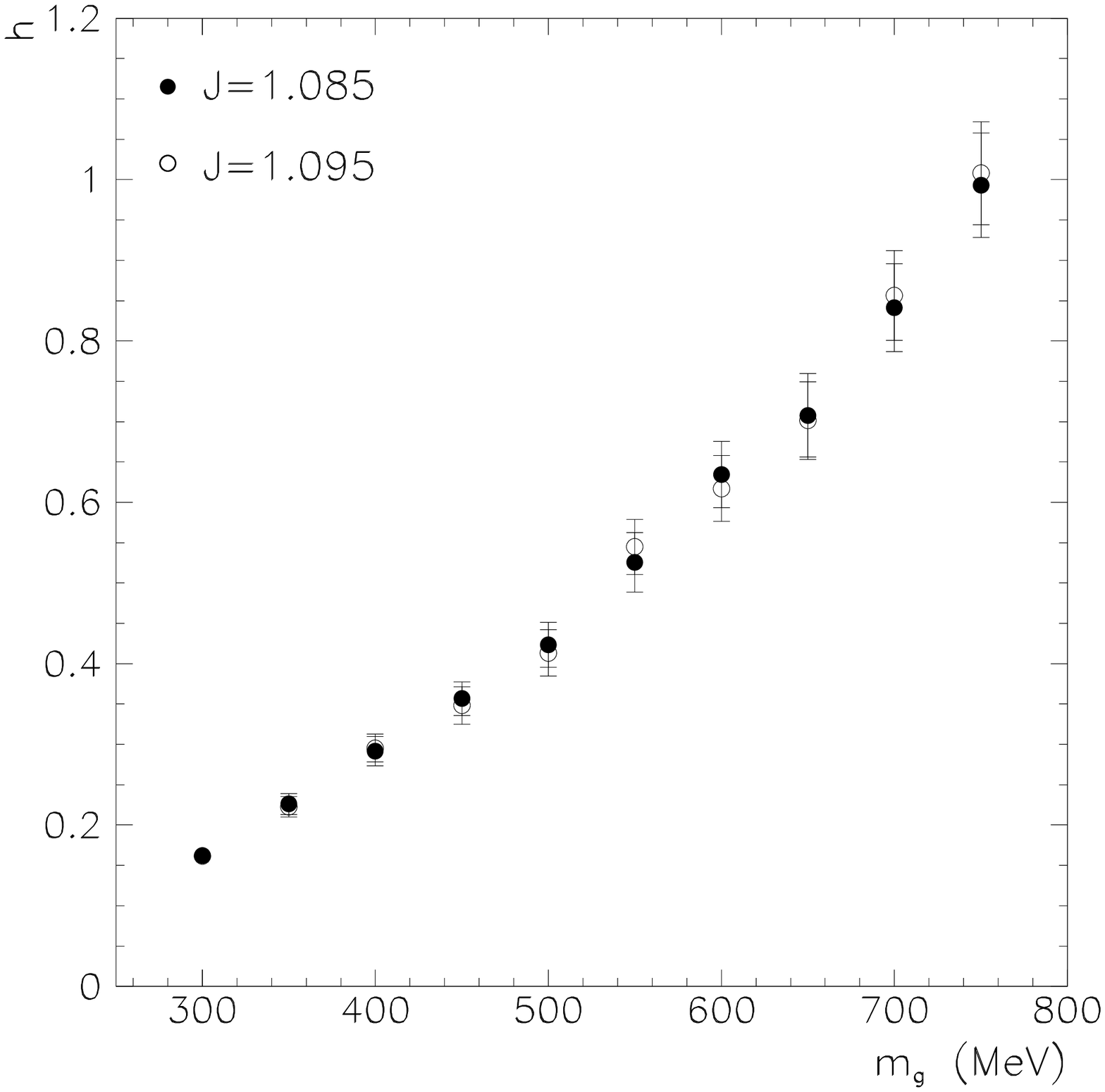}
}
\caption{The factor $h$ as a function of the gluon dynamical mass $m_{g}$ determined by means of the DGM model.}
\label{difdad}
\end{figure}

\begin{figure}
\resizebox{0.97 \textwidth}{!}{%
\includegraphics{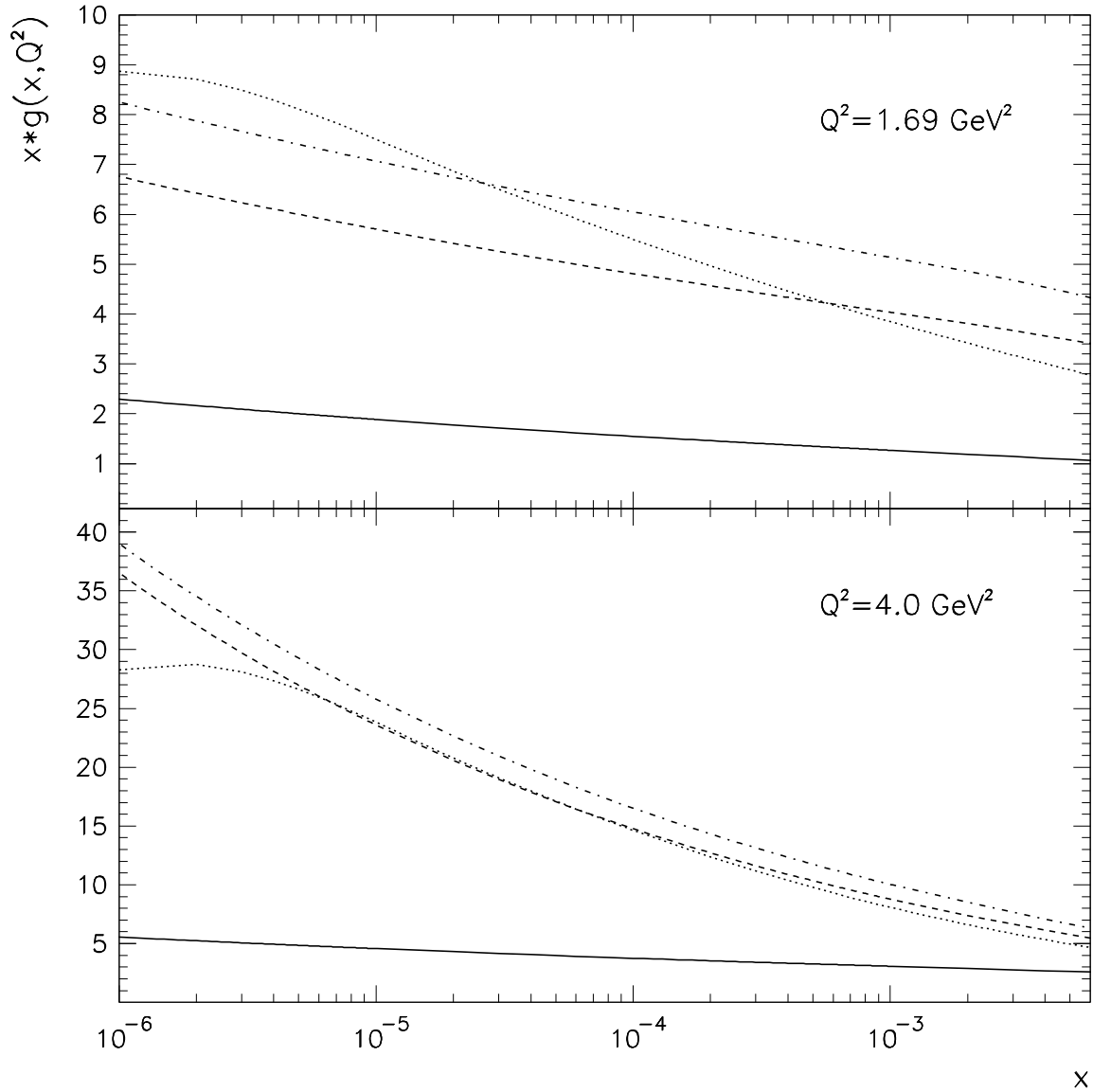}
}
\caption{The gluon distributions at $Q^2=1.69$ and $4.0$ GeV$^2$ from the DGM model (solid curves). The dotted, dashed and
dotted-dashed curves are the distributions corresponding to the MRST, CTEQ6L1 and CTEQ6L gluon sets, respectively.}
\label{difdad3}
\end{figure}

\begin{figure}
\resizebox{0.97 \textwidth}{!}{%
\includegraphics{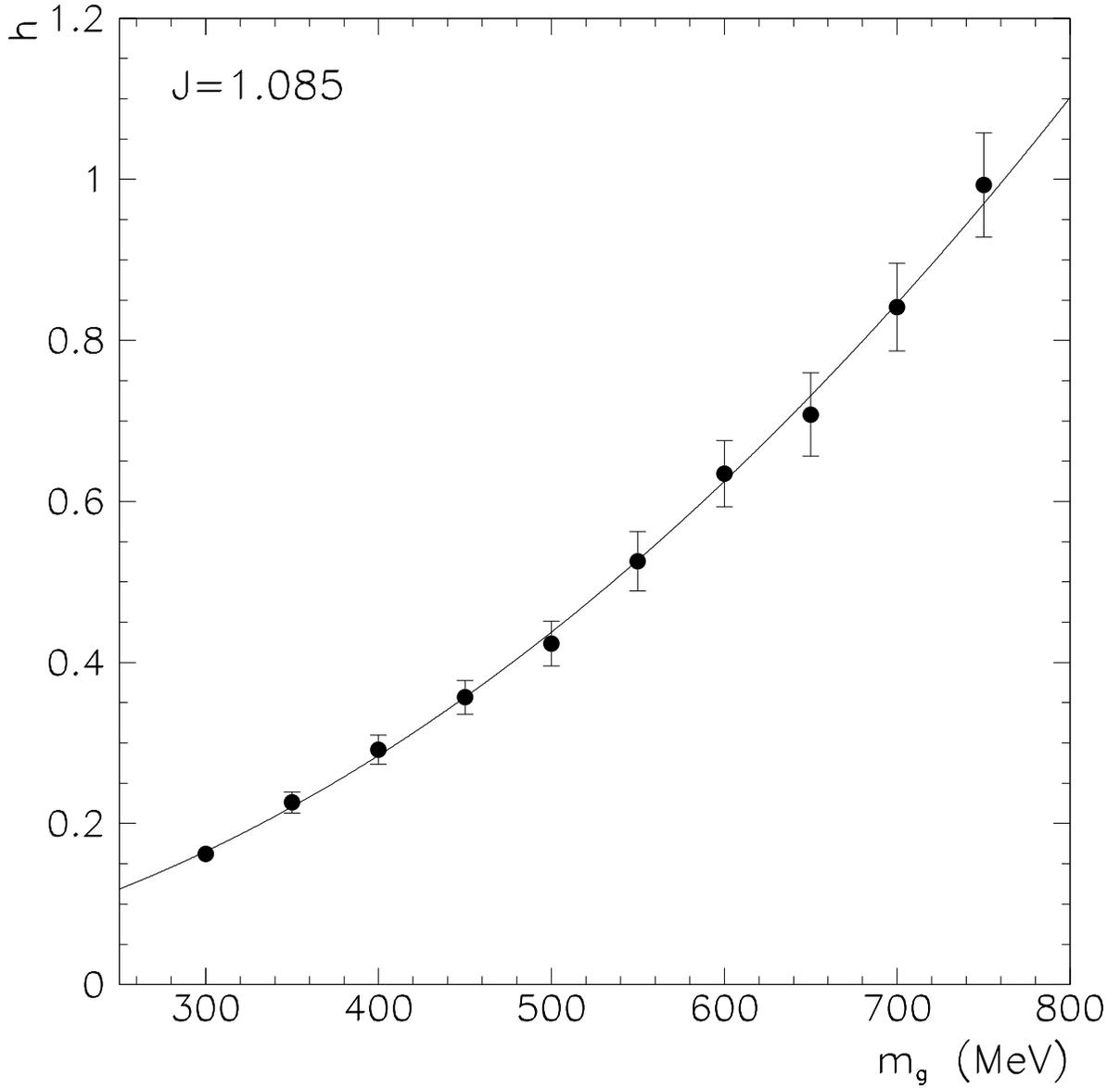}
}
\caption{The factor $h$ as a function of the gluon dynamical mass $m_{g}$. The solid curve is obtained through
the function $h(Q^2)=A+BQ^2$, with $A=0.121\times 10^{-1}$ and $B=0.170\times 10^{-5}$ MeV$^{-2}$.}
\label{difdad2}
\end{figure}

\end{document}